\documentclass[a4paper,11pt]{article}
\usepackage{pos}
\usepackage{booktabs}

\title{New constraints on flavour violating supersymmetry}

\author*[a]{M.A.~Boussejra}
\author[a,b]{F.~Mahmoudi}

\affiliation[a]{Universit\'e de Lyon, Universit\'e Claude Bernard Lyon 1, CNRS/IN2P3, \\
Institut de Physique des 2 Infinis de Lyon, UMR 5822, F-69622, Villeurbanne, France}

\affiliation[b]{Theoretical Physics Department, CERN,
CH-1211 Geneva 23, Switzerland}

\emailAdd{boussejra@ipnl.in2p3.fr}
\emailAdd{nazila@cern.ch}

\abstract{We present a first update on the constraints on general MSSM scenarios with non-minimal sources of flavour violation (NMFV).
Using the public code SuperIso, we present an up-to-date computation of the relevant Wilson coefficients related to $b\to s ll$ transitions which manifest tensions with the SM predictions. We present benchmark scenarios compatible with the most recent fits to the Wilson coefficients. We finally discuss the possibility for such scenarios to completely explain the recent flavour anomalies.
    }

\FullConference{%
  *** The European Physical Society Conference on High Energy Physics (EPS-HEP2021), ***\\
  *** 26-30 July 2021 ***\\
  *** Online conference, jointly organized by Universität Hamburg and the research center DESY ***
}

\begin{document}
\maketitle

\section{Introduction}
In the recent years, the tensions between the Standard Model (SM) and the experimental predictions in the $b \rightarrow s l^+ l^-$ transitions have kept on increasing: a series of measurements have shown $2-3 \sigma$ disagreements with the SM predictions, started with the angular observables (in particular $P_5^\prime$) in the $B^0 \rightarrow K^{*0} \mu^+ \mu^-$ decay (see e.g. \cite{LHCb:2015svh}), followed by the measurement of ratios testing lepton flavour universality \cite{LHCb:2017avl,LHCb:2021trn,LHCb:2021lvy}. The LHCb collaboration has recently measured also the $B^+ \rightarrow K^{*+} \mu^+ \mu^- $ angular observables using the full data coming from the Runs 1 \& 2 corresponding to an integrated luminosity of 9 fb$^{-1}$ \cite{LHCb:2020gog} which confirms the previously observed tensions in the similar neutral decay modes. Model independent global fits to all available $b \rightarrow s l^+ l^-$ data seem to consistently indicate New Physics (NP) compatible with a single shift in $C_9$ from its SM value by about 25\% (see e.g.~\cite{Alguero:2021anc,Hurth:2021nsi,Geng:2021nhg,Bhom:2020lmk}).

In this context of strong and persisting flavour anomalies, the exploration of non-universal flavour models is pertinent to find a  compatible model. In particular, one such promising model is supersymmetry, namely the Minimal Supersymmetric Standard Model (MSSM). 
Until now, most studies conducted in the MSSM assumed harsh constraints on the 105 MSSM free parameters due to computational challenges. Simplifications were made for instance by taking most of the parameters to be zero or assuming the so-called Minimal Flavour Violation (MFV) hypothesis. These approximations of the full MSSM have not proven to be enough to explain the $B$ anomalies so far. Moreover, no signal predicted by such models as the constrained MSSM (cMSSM, 5 free parameters) has been detected at $\sqrt{s} = 13$ TeV at the LHC, or elsewhere. 
Therefore in this study we present a more general approach to $b \rightarrow s l l$ transitions, by looking at the impact of a more general model, namely the phenomenological MSSM (pMSSM, 19 free parameters) by in addition evading the MFV assumptions through the Mass Insertion Approximation (MIA) approach~\cite{Gabrielli:1995bd}.
We then discuss the obtained contributions, and compare them to the pMSSM.

\section{Theoretical Set-up}
\begin{figure}[ht]
        \begin{center}
            \includegraphics[width=0.3\linewidth]{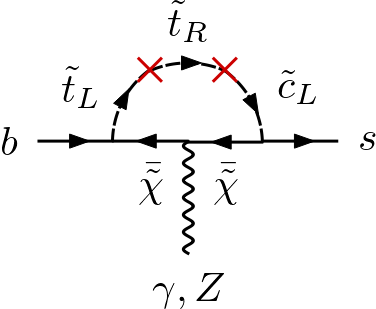}
            \hspace{1cm}
            \includegraphics[width=0.3\linewidth]{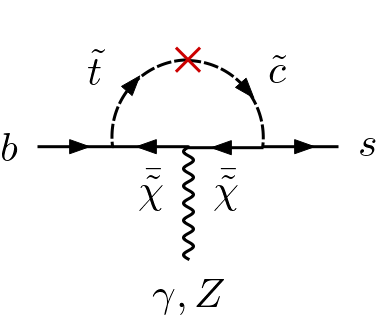}\\
            \vspace{1cm}
            \includegraphics[width=0.3\linewidth]{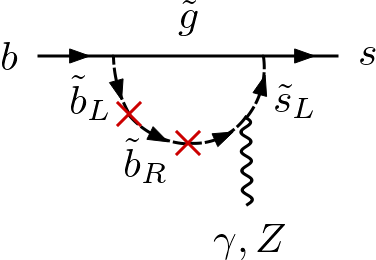}
            \hspace{1cm} \vspace{1cm}
            \includegraphics[width=0.3\linewidth]{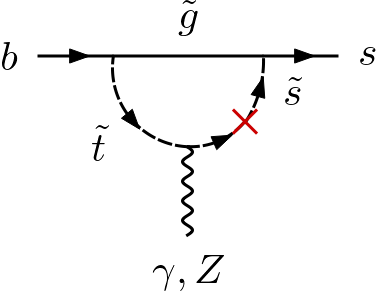}

            \caption[]{Some of the relevant penguin diagrams for $b\rightarrow
            s \ell^+ \ell^-$. The red cross
            indicates a Mass Insertion. Top diagrams are based on chargino interactions.
            The bottom ones consider gluino interactions.
             }
            \protect\label{fig:sfigs}
        \end{center}
    \end{figure}

\begin{figure}[ht]
        \begin{center}
            \includegraphics[scale=0.3]{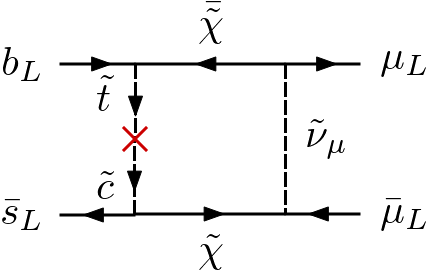}
            \caption[]{Relevant box diagram for $b\rightarrow s \ell^+
            \ell^-$. The red cross
            indicates a Mass Insertion. }
            \protect\label{fig:box}
        \end{center}
    \end{figure}

The pMSSM is known to not be able to shift $C_9$ enough without violating the constraints on $C_7$ coming from $b \rightarrow  s \gamma$ \cite{Mahmoudi:2014mja}. To evade such a limitation, we relax only the MFV hypothesis and use the MIA to turn to Non Minimal Flavour Violating (NMFV) scenarios in this extended pMSSM, by allowing mixing between squarks of the second and third generations. By doing so, additional 1-loop diagrams contribute to $C_9,C_7$ and $C_{10}$, as displayed in Figures~\ref{fig:sfigs},\ref{fig:box}.\\
The MIA gives then a simple way of expressing all relevant quantities~\cite{Dedes:2015twa} (amplitudes, Wilson coefficients, etc.) in term of the flavour violating parameters: in our case the off-diagonal entries of the squark soft-breaking masses. \\
Starting from the squark soft breaking mass matrix and following the conventions of~\cite{Allanach:2008qq}:
\begin{equation}
\label{eq:squark_mass_matrices}
\begin{split}
     \mathcal{M}^2_{\boldsymbol{\tilde{d}}} ~&=~
    \begin{pmatrix}
        M^2_{\tilde{Q}} + m^2_{d} + D_{\tilde{d},L}
        ~~&~~
        \frac{v_d}{\sqrt{2}} T_d^\dag - m_d \mu \tan\beta \\
        \frac{v_d}{\sqrt{2}} T_d - m_d \mu^* \tan\beta  
        ~~&~~
        M^2_{\tilde{D}} + m^2_{d} + D_{\tilde{d},R}
    \end{pmatrix} \, \\ 
    \mathcal{M}^2_{\boldsymbol{\tilde{u}}} ~&=~ 
    \begin{pmatrix}
        V_{\rm CKM} M^2_{\tilde{Q}} V^{\dag}_{\rm CKM} + m^2_{u} + D_{\tilde{u},L}
        ~~&~~
     \frac{v_u}{\sqrt{2}} T_u^\dag - m_u \frac{\mu}{\tan\beta} \\
     \frac{v_u}{\sqrt{2}} T_u - m_u \frac{\mu^*}{\tan\beta}
     ~~&~~
     M^2_{\tilde{U}} + m^2_{u} + D_{\tilde{u},R}
    \end{pmatrix}
    \end{split}
\end{equation}
we define the dimensionless Mass Insertion (MI) parameters as: 
\begin{equation}
    (\delta^{\tilde{f}}_{ij})_{AB} \equiv \frac{(\Delta^{\tilde{f}}_{ij})_{AB}}{M_{sq}}
\end{equation}
where $(\Delta^{\tilde{f}}_{ij})_{AB}$ is an off-diagonal element of the $\tilde{f}=\tilde{u},\tilde{d}$ squark squared mass matrix, while the indices $(i,j) \in \{2,3\}$ span generation space, $(A,B)\in \{L,R\}$ are chirality indices, and $M_{sq}$ is the first and second generations' average squark mass, following the conventions of~\cite{Lunghi:1999uk}. 
    
\section{Results}
We perform a uniform sampling of all the 19 parameters of the pMSSM and the 9 additional MI parameters. We then compute the spectra at the electroweak scale using \texttt{Softsusy}~\cite{Allanach:2001kg} and the pMSSM observables (including the Wilson coefficients) with the publicly available code \texttt{SuperIso}~\cite{Mahmoudi:2008tp}. The additional Wilson coefficient NMFV contributions are computed using the analytical expressions in~\cite{Lunghi:1999uk} including their corrections in~\cite{Behring:2012mv}, and cross-checked with the full calculations performed with \texttt{MARTY}~\cite{Uhlrich:2020ltd}.  

\begin{figure}[ht]
    \centering
    \includegraphics[width=0.7\linewidth]{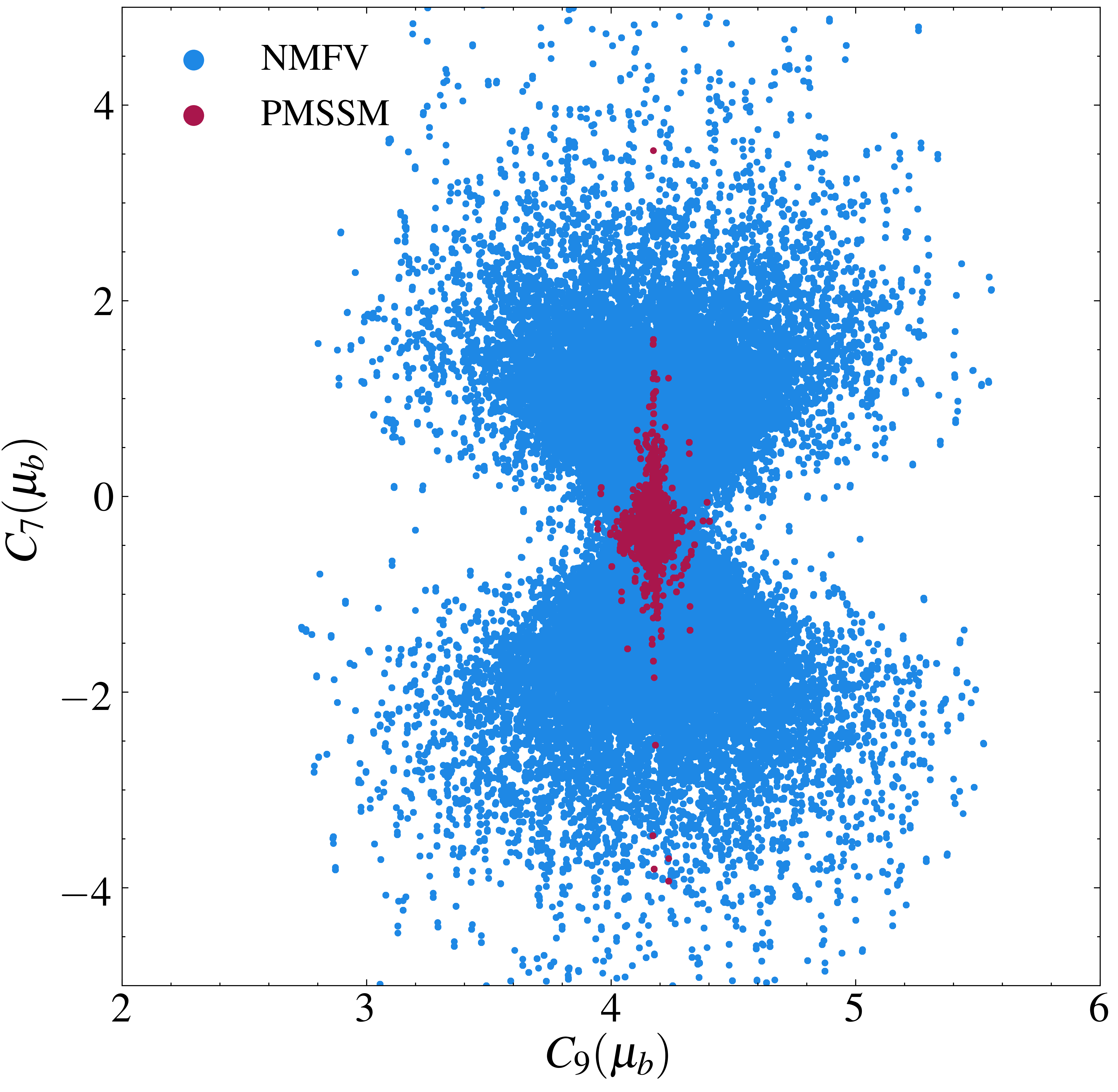}
    \caption{Distribution of the sampled points in the $(C_9,C_7)$ plane. The pMSSM and NMFV distributions are shown in red and blue respectively. Each point in the NMFV is recast to the pMSSM by setting the MIs to zero. In the SM, we have: $C_9^{\rm SM}(\mu_b) = 4.2$.}
    \label{fig:C9C7}
\end{figure}

The results for the Wilson coefficients are then run down to the $m_b$ scale. To better understand the effect of the MI parameters, we also compute the pMSSM Wilson coefficients for the same points by turning the MI parameters to zero. This is shown in Figure~\ref{fig:C9C7}. \\
 As expected, the pMSSM is not able to significantly shift $C_9$, whereas we can see an impressive oyster-shaped spreading of the $C_9,C_7$ distribution in the NMFV case.
 
\begin{figure}[ht!]
    \centering
    \includegraphics[width=0.8\linewidth]{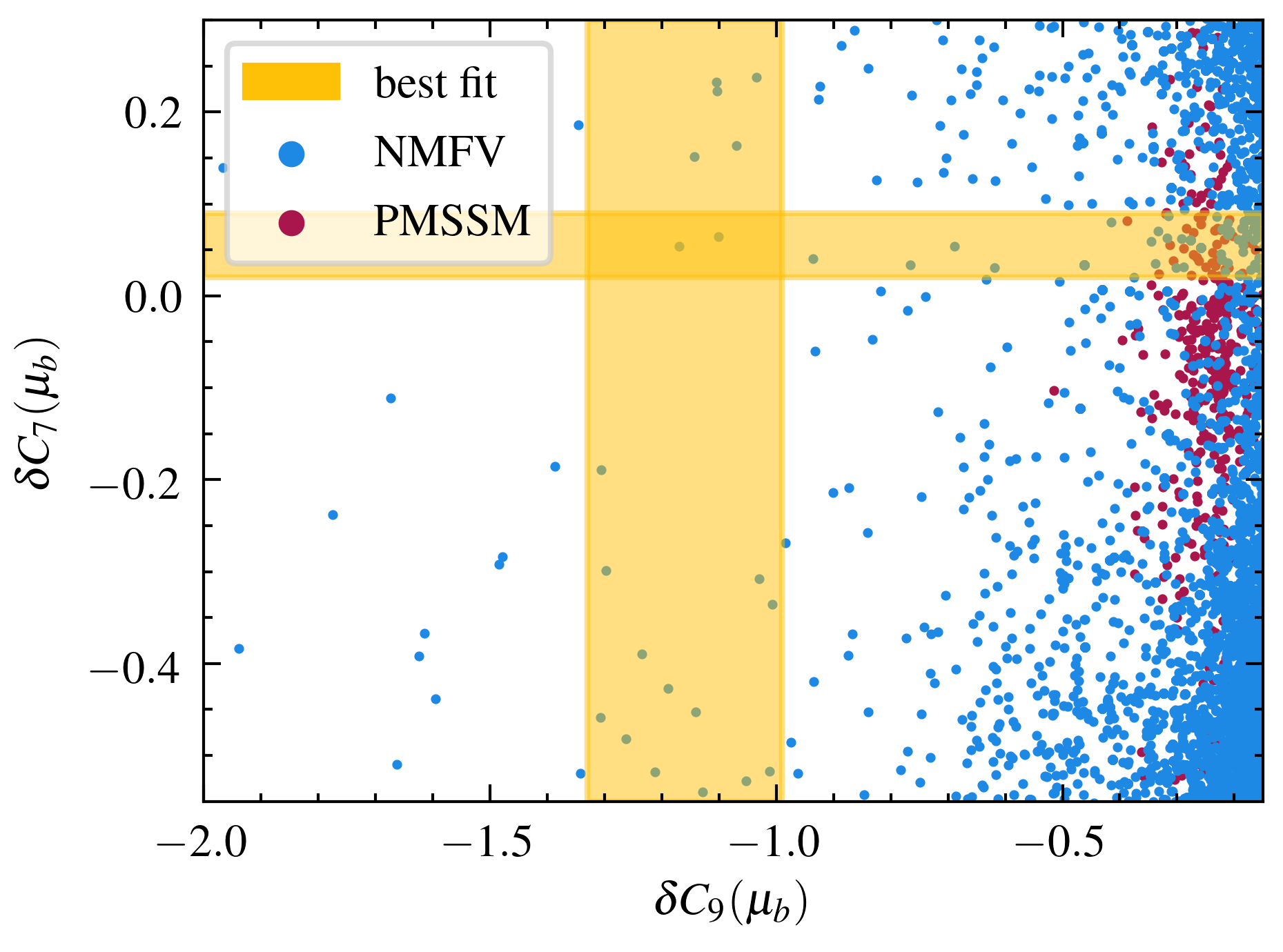}
    \caption{Distribution of the sampled points in the $(\delta C_9, \delta C_7)$ plane, with $\delta C_i = C_i^{\rm NP} - C_i^{\rm SM}$. The best fit regions are shown as orange vertical and horizontal patches.}
    \label{fig:C9bf}
\end{figure}

In Figure~\ref{fig:C9bf}, we calculated the shift from the SM values for $C_9$ and $C_7$, and imposed the theoretical bounds on the MI parameters to ensure no tachyonic squarks are produced in the full spectrum, i.e. all MI parameters should satisfy $ (\delta^{\tilde{f}}_{ij})_{AB} \in [-0.85,0.85]$~\cite{DeCausmaecker:2015yca}. We also impose the LEP constraints on sparticle masses.
The best fit ranges are taken from~\cite{Hurth:2021nsi}. Again, it is clear that the pMSSM fails to fuly account for the $b\rightarrow s l l $ anomalies, while the NMFV is compatible at the 1$\sigma$ level. In Table~\ref{tab:wcf}, we show the Wilson coefficients $C_7, C_9, C_{10}$ for the two best-fit points obtained in our study. These scenarios are compatible with a full explanation of the $B$ anomalies.

Following the recent study of the recast of LHC mass limits when assuming flavour mixing~\cite{Chakraborty:2018rpn}, the corresponding squark and Lightest Supersymmetric Particle (LSP) masses can escape the current mass limits. Also, severely degenerate LSP scenarios as is the case here (with a $\Delta m(\chi^\pm_1,\chi^0_1) \sim0.14 $ GeV ) are an additional experimental challenge for detection in collider events due to extremely soft final states. \\
\begin{table}[ht]
    \centering
    \begin{tabular}{rrr}
\toprule
        $C_7(\mu_b)$ &        $C_9(\mu_b)$ &        $C_{10}(\mu_b)$ \\
\midrule
-0.233 & 3.119 & -3.993 \\
-0.243 & 3.050 & -3.867 \\
\bottomrule
\end{tabular}
    \caption{Wilson coefficient values at the $b$ quark mass scale for the best-fit matching points.}
    \label{tab:wcf}
\end{table}

\section{Conclusion}
We presented a first study of the effect of NMFV scenarios on the pMSSM contributions to the relevant Wilson coefficients for the $b \rightarrow s l l $ anomalies. The NMFV allows to shift these coefficients enough to fully explain the anomalies. Imposing theoretical constraints on the flavour violating parameters leaves compatible benchmark scenarios to further explore.
This clearly shows the interest of NMFV models with respect to more constrained models such as the cMSSM and the pMSSM. These first results are promising, and indicate the necessity of investigating such models in the future, while more and more experimental measurements of lepton flavour violating observables are expected.

\end{document}